# Scaling behavior of fracture properties of tough adhesive hydrogels

Xiang Ni, [†‡] Zhen Yang, [†‡] Jianyu Li[†§*]

† Department of Mechanical Engineering, McGill University, Montreal, QC H3A 0C3, Canada
§ Department of Biomedical Engineering, McGill University, Montreal, QC H3A 0C3, Canada

**ABSTRACT:** Tough adhesive hydrogels find broad applications in engineering and medicine. Such hydrogels feature high resistance against both cohesion and adhesion failure. The superior fracture properties may, however, deteriorate when the hydrogels swell upon exposure of water. The underlying correlation between the polymer fraction and fracture properties of tough adhesive hydrogels remains largely unexplored. Here we study how the cohesion and adhesion energies of a tough adhesive hydrogel evolve with the swelling process. The results reveal a similar scaling law of the two quantities on the polymer fraction ($\phi^y$). Our scaling analysis and computational study reveal that it stems from the scaling of shear modulus. The study will promote the investigation of scaling of hydrogel fracture and provide development guidelines for new tough adhesive hydrogels.

Tough adhesive hydrogels are highly resistant to crack growth within the matrix, and capable of adhering strongly to diverse substrates such as biological tissues and elastomers[1, 2]. The tough adhesive hydrogels are in contrast to conventional hydrogels, which are vulnerable to cohesion failure (i.e., fracture in the bulk) and adhesion failure (i.e., debonding at the interface). Thanks to the exceptional properties, tough adhesive hydrogels open a plethora of applications such as tissue adhesives[3, 4], wound dressings[5], soft robotics[6, 7], and implantable devices[8, 9]. In many applications, the hydrogels are exposed to water and readily swell. The resulting change of the polymer fraction (i.e., the water content) may significantly alter the fracture properties of the hydrogels[10]. The underlying correlation is thus fundamental and practically important to explore.

The coupling between the swelling and fracture of tough adhesive hydrogels poses challenges to theoretical analysis[11, 12], and computational modeling[13, 14]. It is in part because both the cohesion and adhesion failure of tough adhesive hydrogels involve stress/strain fields of highly nonlinear nature and complex physicochemical interactions. To this end, the scaling theory by de Gennes[15] and many others is appealing. Because it can capture key structure-property relations, such as the dependence of elastic modulus and polymer content, to reveal the fundamentals and enable material design. However, few reports have been seen on the scaling of fracture properties (e.g., cohesion and adhesion energies). One exception is a recent work reporting the elastic modulus and the intrinsic cohesion energy (i.e., intrinsic toughness with no background hysteresis involved) of polyacrylamide hydrogels scales the same way as a function of the polymer fraction[16]. In the case of tough adhesive hydrogels, the fracture properties (either cohesion or adhesion) are dominated by background hysteresis, whereas the intrinsic toughness plays a minor role (Figure S1)[17]. Little is known about the scaling of the polymer fraction and the cohesion and adhesion energies, which is the focus of the current study. We will swell the tough adhesive hydrogels for varying polymer fractions and then characterize their cohesion and adhesion energies, respectively. The results will be further analyzed with a scaling theory developed by Brown[18] and finite element simulation. This work would lead to insights and experimental methods to control and predict the mechanics of tough adhesive hydrogels.

In this study, we choose an alginate-polyacrylamide hydrogel as the model hydrogel and porcine skin as the model substrate[19]. The hydrogel is extremely tough and can form tough adhesion on various substrates, including hydrogels and tissues, using recently developed strategies[1, 2, 20]. We synthesized and swelled the alginate-polyacrylamide hydrogels in PBS for different durations (0-6 hours). The choice of using PBS is based on broad biomedical applications of tissue adhesive hydrogels, since it better mimics the physiological environment. Note that in addition to the change of the polymer fraction as the hydrogel swells, the crosslink density of the first network was changed as well, due to the ion change between the hydrogel and the PBS solution. The resulting hydrogels with varying polymer fractions were then placed in a sealed bag for a homogeneous water distribution inside (Figure 1a). We then activated the hydrogel surface with chitosan and EDC/NHS and compressed the hydrogel to porcine skin to form tough adhesion[1]. We measured the weight of the wet gel and that of solid content after lyophilization, and then calculated the polymer fraction $\phi$ (See Experimental Methods). With longer swelling, the hydrogels absorb more water and further lower the polymer fraction (Figure S2), which can be varied between 5% and 10%. In addition to the change of the polymer fraction, the nominal crosslink density of the Alginate-$Ca^{2+}$ was also progressively dissociated upon swelling, due to the ion exchange between the hydrogel and PBS.

Next, we characterize the cohesion energy and adhesion energy of the swollen hydrogels using tear and 180-degree soft peeling methods, respectively. The tear testing has been widely used to measure cohesion energy (i.e., toughness) of materials such as hydrogels[21] and elastomers[22], and is chosen in this study to ensure crack propagation through the bulk matrix. While the arms of the specimen are pulled unidirectionally at a constant speed, the pulling force increases to a plateau (Figure 1b-d). The averaged plateau force F is used to calculate the cohesion energy via $\Gamma_C = 2F/t$. As expected, the cohesion energy decreases as the hydrogel matrix swells. We also characterize the adhesion energy between the tough adhesive hydrogel and the porcine skin with 180-degree peeling tests (Figure 1e-g). From the peeling tests, the plateau peeling force is used to calculate the adhesion energy via $\Gamma_A = 2F/w$. Similar to the cohesion energy, the adhesion energy exhibits a positive correlation with the polymer fraction. Interestingly, when the polymer fraction is below 5%, no tough adhesion is formed. Because the adhesion is so weak that the interface debonds well before the two arms of the specimen are stretched to 180 degrees. The phenomenon can be understood that the large mesh size of the swollen hydrogel exceeds the limit for the formation of the entanglements between the chitosan chains and the hydrogel network at the interface, which underpins the tough hydrogel adhesion. As a result, poor adhesion was observed in the highly swollen hydrogel. A possible method to solve this problem is using chemical bonding rather than bridging polymer as proposed by Yuk, Zhang et al[2]. Together, the results show that modulating the polymer fraction can tune cohesion and adhesion energies of tough adhesive hydrogels over a wide range, and that there is a positive correlation between the polymer fraction and the fracture properties.

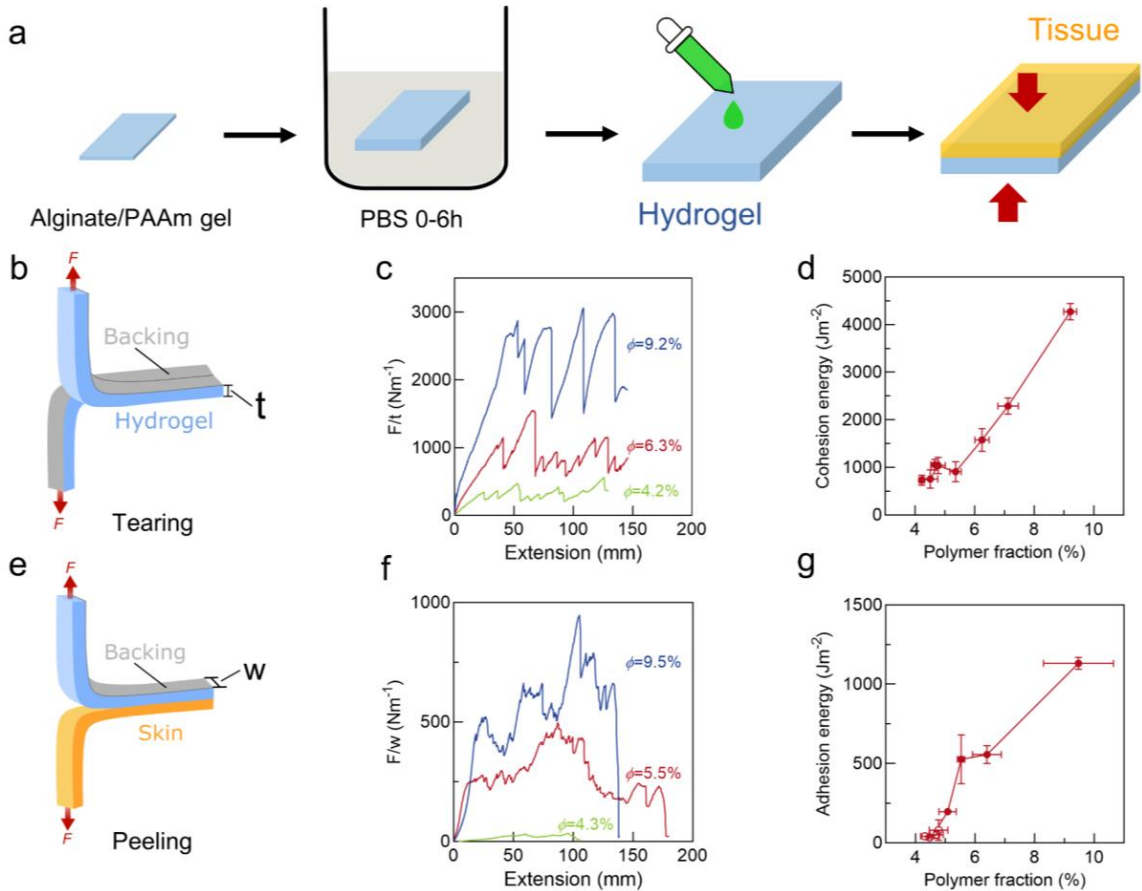

**Figure 1:** Fracture measurements of tissue adhesive hydrogels with varying polymer fractions. (a) As prepared alginate/PAAm hydrogels were soaked in PBS solutions for varying durations, then activated with chitosan and coupling agents to form adhesion with tissues under compression. (b-d) Cohesion energy measurement with tearing test. (b) A specimen of thickness t is pulled vertically with a constant speed. (c) Tearing force per thickness of hydrogels as a function of extension and polymer fraction. (d) Cohesion energy as a function of polymer fraction. (e-g) Adhesion energy measurement with 180-degree peeling test. The specimen width is w. (f) Peeling force per width as a function of extension and polymer fraction. (g) Adhesion energy as a function of polymer fraction, empty circles indicate weak bonding. Sample size N=3.

We then study the scaling of the cohesion and adhesion energies as a function of the polymer fraction. Figure 2 plots the preceding results in a log-log scale, showing an exponent of 2.2 and an identical scaling law ($\phi^{2.2}$) of the cohesion and adhesion energies. First, we hypothesize the scaling relation is attributable to the dependence of elastic modulus on the polymer fraction. To test the hypothesis, we perform a scaling analysis based on Brown's model for fracture of double network hydrogels[18]. This model can be applied to the tough hydrogel in this study since the key assumption of Brown's model is met: the elastic modulus at loading stage before yielding $E_1$ is much larger than that at unloading stage $E_2$ (Fig.S3). In adapting Brown's model (Figure 2a) to tough hydrogel, it is worth noting that the 1st network (alginate) and the 2nd (polyacrylamide) both contribute to the initial regime of the stress-strain curve (slope $E_1$), but the former mainly takes up since the 1st network (alginate) is much stiffer and more brittle than the 2nd network (polyacrylamide). When the strain reaches a critical value $\varepsilon_1$, the 1st network undergoes progressive damage, as manifested by a yielding regime at a critical stress $\sigma_1 = E_1\varepsilon_1$ (region 2). When the sample is further loaded to the maximum strain $\varepsilon_2$, the 1st network damages completely, and the 2nd network bridges the crack and bears all the load. As such, the stress-strain curve undergoes an abrupt decrease to a stress level $\sigma_2 = E_2\varepsilon_2$ (region 3). After that, the 2nd network dominates the unloading response with a slope $E_2$ (region 4). This model allows us to relate the cohesion energy (i.e., fracture energy) to the elastic modulus without invoking complex fields as follows.

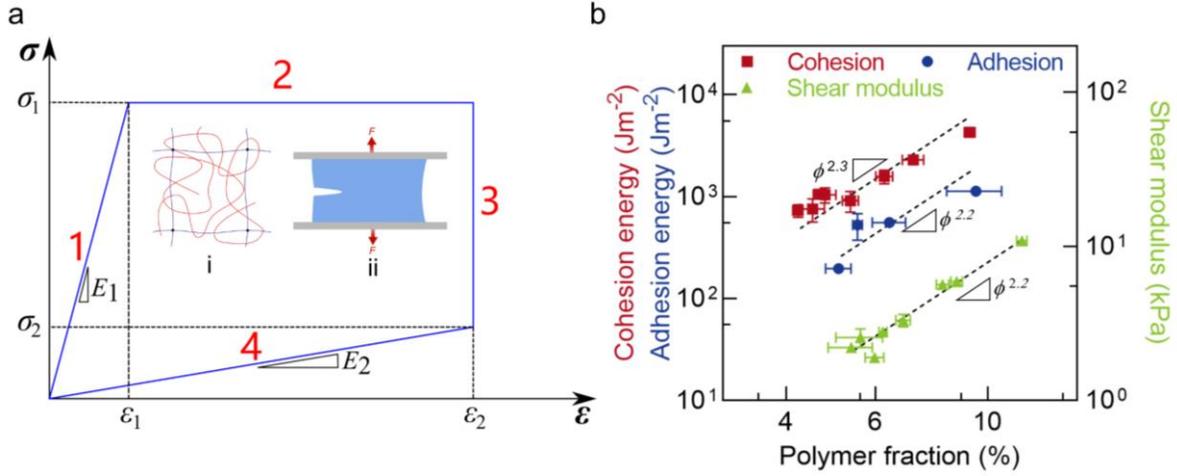

**Figure 2:** Scaling relations between mechanical properties and polymer fraction. (a) Tensile stress-strain curve of a loading-unloading cycle in Brown's model divided into 4 parts. Inserted figures: i) Illustration of structure of double network hydrogel. ii) Pure shear configuration. (b) Log-log plot illustrating the cohesion and adhesion energies, shear modulus as a function of polymer fraction. Sample size N=3.

Consider a steady crack propagation, within the damage zone (which resembles a sub-pure shear specimen), the first stiff network is assumed to be completely damaged and the material response is dominated by the elastic modulus $E_2$ of the 2nd network. By leveraging the energy release rate expression for the pure shear specimen, $G_{tip} = Wh$, and the critical condition for steady crack propagation $G_{tip} = \Gamma_{Co}$, the intrinsic cohesion energy can be approximated as $\Gamma_{Co} \approx E_2\varepsilon_2^2 h$, where $h$ is height of the sub-pure shear specimen. On the other hand, the apparent cohesion energy equals the energy per unit length required to expand the damage zone from 0 to $h$ in the undeformed configuration, thus can approximated by $G_{apparent} = \Gamma_C \approx \sigma_1 h \varepsilon_2$. Note in the above expression of the apparent and intrinsic cohesion energies, the numerical pre-factors have been dropped. To compare with the experimental results in Figure 2, we recast the expression of the cohesion energy into the form $\Gamma_C \approx \Gamma_{Co}(E_1\varepsilon_1)/(E_2\varepsilon_2)$.

It is reasonable to assume that the critical strain $\varepsilon_1$ corresponding to the yield stress of the 1st network and the maximum strain at which the same network is completely broken $\varepsilon_2$ have the same scaling with respect to the polymer fraction. Since the 2nd network is essentially polyacrylamide hydrogel as investigated by Li et al[16], we adopt their finding that the intrinsic cohesion energy $\Gamma_{Co}$ and the elastic modulus $E_2$ of the polyacrylamide network share the same scaling with respect to the polymer fraction. Also, we found that the scaling exponent between intrinsic cohesion energy $\Gamma_{Co}$ and polymer fraction is 0.76 (Figure. S4) which is close to that of the elastic modulus of PAAm hydrogel reported by Li et al[16]. Thus, we assume the $\Gamma_{Co} \sim E_2$ relationship of tough gels matches that of the PAAm gels. It should be noted that the assumption calls for experimental validation in future work. As the shear modulus of the tough hydrogel $\mu \sim E_1$, we conjecture the cohesion energy and the shear modulus share the same scaling as a function of polymer content, $\Gamma_C \sim \mu$.

To testify the hypothesis, we then characterize the shear modulus $\mu$ of the tough adhesive hydrogel using a rheometer (TA HR-2 hybrid rheometer). The swollen hydrogels are gently compressed between parallel plates on the rheometer and subject to a shear strain of 1%. As expected, the shear modulus decreases with the polymer fraction. Remarkably, the result reveals the same scaling relation ($\phi^{2.2}$) between the shear modulus and the polymer fraction (Figure 2b). The interpretation of the exponent 2.2 requires further investigation, as different scaling relations have been reported for hydrogels with different solvent contents and network configuration[23-25]. To support our results that the cohesion energy of the tough hydrogel has the same scaling to that of the shear modulus, we also measured toughness and shear modulus of PNIPAm/Alginate-Ca$^{2+}$ hydrogel at various polymer fraction. It shows that the scaling for toughness ($\phi^{1.8}$) is close to that for shear modulus ($\phi^{1.6}$). The result confirms the proceeding hypothesis and supports the scaling analysis based on Brown's model for tough hydrogels.

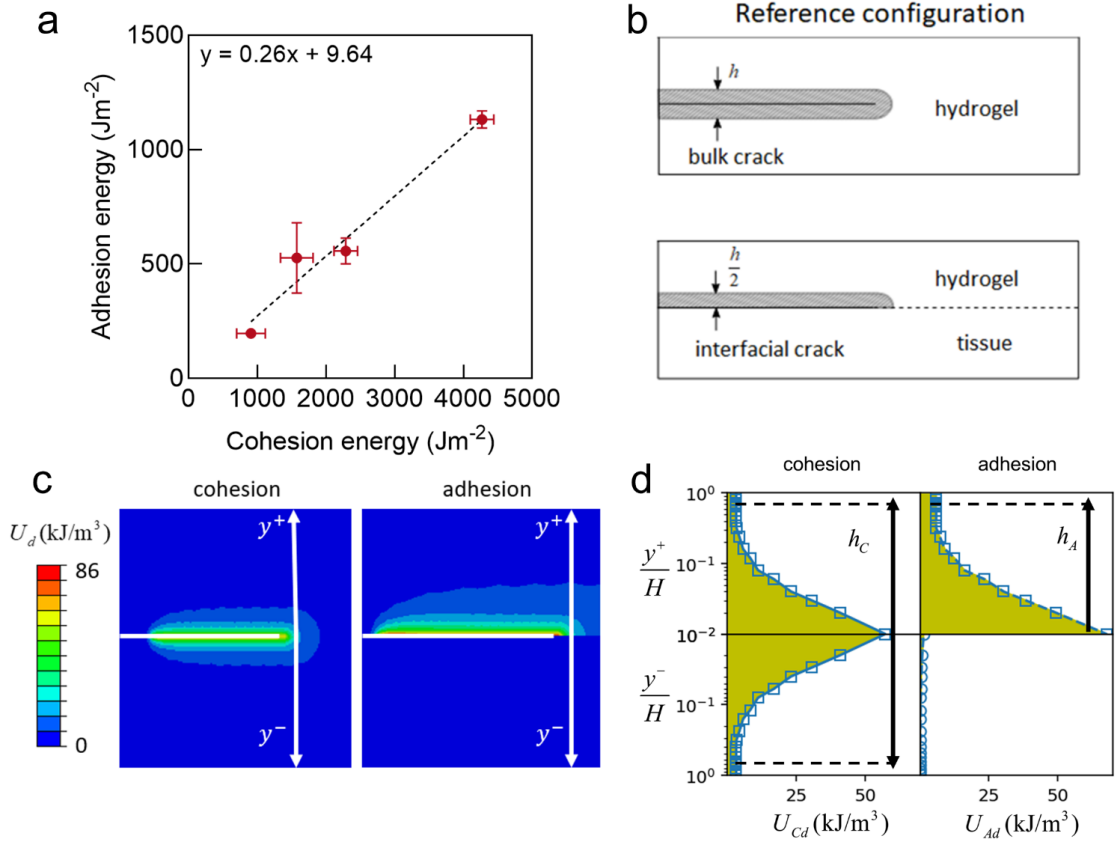

**Figure 3:** Relationship between adhesion energy and cohesion energy. (a) Adhesion energy is about one quarter of cohesion energy. Sample size N=3. (b) Damages zones near a bulk crack and an interfacial crack in the reference configuration of a pure shear specimen during the steady-state crack propagation. (c) A similar plot to (b) with contour plots showing the distribution of the dissipated energy per unit volume $U_d$. The white horizontal lines denote the initial cracks. (d) plots the $U_d$ distribution across the height extracted from the vertical axes in (c). The colored area enclosed by the $U_d$ curve and the y-axis denotes the total dissipated energy in the specimen of height 2H.

We further perform a finite element (FE) study to delineate the correlation between the cohesion and adhesion energies (Figure 3). Our results reveal a simple relation between the two fracture properties (Figure 3a). The ratio of adhesion and cohesion energies can be formulated by $\frac{\Gamma_A}{\Gamma_C} \approx \frac{\Gamma_{Ao}}{\Gamma_{Co}} \frac{\xi_A}{\xi_C}$, where $\Gamma_{io}$ is the intrinsic toughness and $\xi_i$ is the energy enhancement factor due to background hysteresis; the subscription $i$ can be $C$ or $A$, referring to cohesion and adhesion, respectively. The decoupling has been validated both computationally and analytically[12, 13]. We also adopt the FE model by Zhang et al.[13] to simulate the fracture of a pure shear specimen (See the Supporting Information for the Method). We find that the energy enhancement $\xi$ is indeed independent of the intrinsic toughness $\Gamma_o$, indicating that the apparent fracture energy has a separable dependence on $\Gamma_{io}$ and $\xi$.

For the ratio of energy enhancement factors $\xi_A / \xi_C$, we resort to the FE simulation. For convenience, the intrinsic adhesion energy is set to be equal to the intrinsic cohesion energy. We hypothesize that the region of background hysteresis of a crack into the bulk hydrogel (cohesion) is approximately two times that of an interfacial crack between the hydrogel and the skin (adhesion), as the skin is much stiffer thus contributing to negligible dissipation (Figure 3b). For the steady crack propagation at the critical stretch $\lambda_c$, the cohesion and adhesion energies of the tough adhesive hydrogels can be approximated by the energy dissipated by the hydrogel and the tissue. The dissipated energy can be determined by integrating the maximum dissipated energy of a unit volume ahead of the crack tip in its loading history $U_d$ over the height of the specimen (2H)[12]. With the same intrinsic toughness, we have

$\xi_A / \xi_C = \int_{-H}^{H} U_{Ad}(y)dy / \int_{-H}^{H} U_{Cd}(y)dy$. As can be seen in Figure 3c, in the cohesion case the upper and the lower parts contribute equally to the total energy dissipation, whereas in the adhesion case only the upper part (hydrogel adhesive) contributes to the energy dissipation and the lower part (tissue) shows negligible energy dissipation. Figure 3d plots the energy dissipation distributions across the height in the adhesion and the cohesion cases. Upon the observation, the $U_{Ad}$ and $U_{Cd}$ distributions in the upper parts show a qualitatively similar trend. By adopting a customized definition of the size of the dissipation zone within which 90% of the total energy dissipation in the hydrogel adhesive is manifested, we have $h_A = \frac{1}{2} h_C \approx 0.8H$, which confirms our hypothesis. The ratio of energy enhancement factors $\xi_A / \xi_C$, however, is not exactly $\frac{1}{2}$, depending on a dimensionless parameter $S_{\text{interface}} / \mu$, with $S_{\text{interface}}$ being the maximum cohesive strength. This is likely due to the different crack tip fields in the pure-shear and the interfacial crack specimen, which is not the focus of our current study. Nevertheless, the ratio $\xi_A / \xi_C$ was found to flip around 0.7 for a variety choice of $S_{\text{interface}} / \mu$. The intrinsic cohesion and adhesion energies of the as-prepared tough hydrogel we measured were 62.7 Jm$^{-2}$ and 26.4 Jm$^{-2}$ (Figure. S7), respectively, which are close to the data reported in literature. Altogether, the ratio of adhesion and cohesion energies is calculated as 0.3, in a good agreement with the experimental results.

In summary, we study the coupling between the swelling and fracture behavior of tough adhesive hydrogels. Our results reveal that both cohesion and adhesion energies of tough adhesive hydrogels decay with the swelling process. Thanks to the unique toughening nature of the tough adhesive hydrogel, a simple scaling law exists for the cohesion and adhesion energies as well as the shear modulus to the polymer fraction. The scaling law is corroborated with the scaling analysis and finite element simulation. It is worth noting that the findings are limited to tough adhesive hydrogels whose fracture behavior is governed by background hysteresis. Besides, the scaling value may vary in different solution environments, but our conclusion that $\Gamma_A \sim \Gamma_C \sim \mu$ still retains. This study provides a facile approach to control and predict the fracture properties of hydrogels. It also calls for further development in the scaling theory of tough hydrogels as they gain increasing impacts in broad areas.

## Author Contributions

‡ These authors contributed equally. X. N. performed experiments and Z. Y. conducted analysis and simulation. J. L. supervised the project. All authors contributed to and approved the manuscript.

## REFERENCES

1. Li, J.; Celiz, A. D.; Yang, J.; Yang, Q.; Wamala, I.; Whyte, W.; Seo, B. R.; Vasilyev, N. V.; Vlassak, J. J.; Suo, Z.; Mooney, D. J., Tough Adhesives for Diverse Wet Surfaces. *Science.* **2017,** *357* (6349), 378-381.
2. Yuk, H.; Zhang, T.; Lin, S.; Parada, G. A.; Zhao, X., Tough Bonding of Hydrogels to Diverse Non-Porous Surfaces. *Nature materials.* **2016,** *15* (2), 190.
3. Annabi, N.; Zhang, Y.-N.; Assmann, A.; Sani, E. S.; Cheng, G.; Lassaletta, A. D.; Vegh, A.; Dehghani, B.; Ruiz-Esparza, G. U.; Wang, X., Engineering a Highly Elastic Human Protein–Based Sealant for Surgical Applications. *Sci Transl Med.* **2017,** *9* (410), eaai7466.
4. Mehdizadeh, M.; Yang, J., Design Strategies and Applications of Tissue Bioadhesives. *Macromol Biosci.* **2013,** *13* (3), 271-288.
5. Blacklow, S. O.; Li, J.; Freedman, B. R.; Zeidi, M.; Chen, C.; Mooney, D. J., Bioinspired Mechanically Active Adhesive Dressings to Accelerate Wound Closure. *Sci Adv.* **2019,** *5* (7), eaaw3963.
6. Kim, Y.; Parada, G. A.; Liu, S.; Zhao, X., Ferromagnetic Soft Continuum Robots. *Science Robotics.* **2019,** *4* (33), eaax7329.
7. Liu, X.; Liu, J.; Lin, S.; Zhao, X., Hydrogel Machines. *Materials Today.* **2020**.
8. Seliktar, D., Designing Cell-Compatible Hydrogels for Biomedical Applications. *Science.* **2012,** *336* (6085), 1124-1128.
9. Zhang, L.; Cao, Z.; Bai, T.; Carr, L.; Ella-Menye, J.-R.; Irvin, C.; Ratner, B. D.; Jiang, S., Zwitterionic Hydrogels Implanted in Mice Resist the Foreign-Body Reaction. *Nature Biotechnology.* **2013,** *31* (6), 553-556.
10. Sato, K.; Nakajima, T.; Hisamatsu, T.; Nonoyama, T.; Kurokawa, T.; Gong, J. P., Phase-Separation-Induced Anomalous Stiffening, Toughening, and Self-Healing of Polyacrylamide Gels. *Adv Mater.* **2015,** *27* (43), 6990-6998.
11. Long, R.; Hui, C.-Y., Fracture Toughness of Hydrogels: Measurement and Interpretation. *Soft Matter.* **2016,** *12* (39), 8069-8086.
12. Qi, Y.; Caillard, J.; Long, R., Fracture Toughness of Soft Materials with Rate-Independent Hysteresis. *Journal of the Mechanics and Physics of Solids.* **2018,** *118*, 341-364.
13. Zhang, T.; Lin, S.; Yuk, H.; Zhao, X., Predicting Fracture Energies and Crack-Tip Fields of Soft Tough Materials. *Extreme Mech Lett.* **2015,** *4*, 1-8.
14. Zhang, T.; Yuk, H.; Lin, S.; Parada, G. A.; Zhao, X., Tough and Tunable Adhesion of Hydrogels: Experiments and Models. *Acta Mechanica Sinica.* **2017,** *33* (3), 543-554.
15. De Gennes, P.-G.; Gennes, P.-G., *Scaling Concepts in Polymer Physics*. Cornell university press: 1979.
16. Li, Z.; Liu, Z.; Ng, T. Y.; Sharma, P., The Effect of Water Content on the Elastic Modulus and Fracture Energy of Hydrogel. *Extreme Mech Lett.* **2019**.

17. Zhao, X., Designing Toughness and Strength for Soft Materials. *Proceedings of the National Academy of Sciences.* **2017,** *114* (31), 8138-8140.
18. Brown, H. R., A Model of the Fracture of Double Network Gels. *Macromolecules.* **2007,** *40* (10), 3815-3818.
19. Sun, J.-Y.; Zhao, X.; Illeperuma, W. R.; Chaudhuri, O.; Oh, K. H.; Mooney, D. J.; Vlassak, J. J.; Suo, Z., Highly Stretchable and Tough Hydrogels. *Nature.* **2012,** *489* (7414), 133.
20. Yang, J. W.; Bai, R. B.; Suo, Z. G., Topological Adhesion of Wet Materials. *Adv Mater.* **2018,** *30* (25).
21. Bai, R.; Chen, B.; Yang, J.; Suo, Z., Tearing a Hydrogel of Complex Rheology. *Journal of the Mechanics and Physics of Solids.* **2019,** *125*, 749-761.
22. Gent, A. N., Adhesion and Strength of Viscoelastic Solids. Is There a Relationship between Adhesion and Bulk Properties? *Langmuir.* **1996,** *12* (19), 4492-4496.
23. Obukhov, S. P.; Rubinstein, M.; Colby, R. H., Network Modulus and Superelasticity. *Macromolecules.* **1994,** *27* (12), 3191-3198.
24. Hoshino, K.-i.; Nakajima, T.; Matsuda, T.; Sakai, T.; Gong, J. P., Network Elasticity of a Model Hydrogel as a Function of Swelling Ratio: From Shrinking to Extreme Swelling States. *Soft Matter.* **2018,** *14* (47), 9693-9701.
25. Sakai, T.; Kurakazu, M.; Akagi, Y.; Shibayama, M.; Chung, U.-i., Effect of Swelling and Deswelling on the Elasticity of Polymer Networks in the Dilute to Semi-Dilute Region. *Soft Matter.* **2012,** *8* (9), 2730-2736.